\documentclass[aps,prd,preprint,nofootinbib]{revtex4}
\usepackage{mathrsfs}
\usepackage{amsfonts}
\usepackage{epsfig}
\usepackage{graphicx}
\newcommand{\bqa}{\begin{eqnarray}}
\newcommand{\eqa}{\end{eqnarray}}

\begin{document}
\title{The Next-to-Leading Order Corrections to
Top Quark Decays to Heavy Quarkonia \\[9mm]}

\author{Peng Sun$^{1}$, Li-Ping Sun$^{1}$, Cong-Feng Qiao$^{1,2}
\footnote{corresponding author}$ } \affiliation{$^{1}$College of
Physical Sciences, Graduate University of Chinese Academy of
Sciences \\ YuQuan Road 19A, Beijing 100049, China}
\affiliation{$^{2}$Theoretical Physics Center for Science Facilities
(TPCSF), CAS\\ YuQuan Road 19B, Beijing 100049, China}

\author{~\vspace{0.9cm}}

\begin{abstract}
\vspace{3mm} The decay widths of top quark to S-wave $b\bar{c}$ and
$b\bar{b}$ bound states are evaluated at the next-to-leading(NLO)
accuracy in strong interaction. Numerical calculation shows that the
NLO corrections to these processes are remarkable. The quantum
chromodynamics(QCD) renormalization scale dependence of the results
is obviously depressed, and hence the uncertainties lying in the
leading order calculation are reduced.

\vspace {7mm} \noindent {\bf PACS number(s):} 13.85.Ni, 14.40.Lb,
12.39.Hg, 12.38.Bx.

\end{abstract}

\maketitle
\section{Introduction}
Since predicted by the Standard Model (SM)\cite{top3,top4,top5}, top
quark has become an important role in high energy physics due to its
large mass, which is close to the electroweak symmetry breaking
scale \cite{top1}. A great deal of researches focusing on top quark
physics have been performed after its discovery in 1995 in the
Fermilab \cite{top6}. On the experiment aspect, with the running of
the Tevatron and forthcoming LHC, the lack of adequate events will
not be an obstacle for the top quark physics study. According to
Ref. \cite{top2}, at the LHC $10^{7}\sim10^{8}$ $t\bar{t}$ pairs can
be obtained per year, so this enables people to measure various top
quark decay channels. Meanwhile, the copious production of the top
quarks supplies also a great number of bottom quark mesons since the
dominant top quark decay channel is $t\rightarrow b + W^+$.
Therefore, the bottom quark meson production in top quark decays may
stand as an important and independent means for the study of heavy
meson physics and the test of perturbative QCD (pQCD).

As the known heaviest mesons, bottomonia and $B_c$ ($b\bar{c}$ or
$\bar{b}c$) possess particular meaning in the study of heavy flavor
physics. The LHCb as a detector specifically for the heavy flavor
study at the LHC will supply copious $B_c$ and $\Upsilon$ data for
this aim. Theoretically, the direct hadroproduction of $B_c$ and
$\Upsilon$ was studied in the literature \cite{Bc1,Bc7,Bs7}. In
addition to the ``direct'' production, ``indirect'' process as in
top quark decays may stand as an independent and important source
for $B_c$ and $\Upsilon$ production. Since the top quark's lifetime
is too short to form a bound state \cite{top7}, the $B_c$ and
$\Upsilon$ production involved scheme in top quark decays is less
affected by the non-pertubative effects than in other processes. In
Ref. \cite{Bc2}, the top quark decays into $\Upsilon$ and
${\bar{B}_c}^{*}$ at the Born level was evaluated. Recently, the S-
and P-wave $B_c$ meson productions in top quark decays were fully
evaluated, including the color-octet contributions, at the leading
order accuracy of QCD by Chang $et$ $al.$ \cite{Bc3}.

Considering the importance of investigating $B_c$ and $\Upsilon$ in
the study of perturbative Quantum Chromodynamics (pQCD) and
potential model, it is reasonable and interesting to evaluate the
production rates of these mesons in top quark decays at the
next-to-leading order (NLO) accuracy of pQCD. At the bottom quark
and charm quark mass scales the strong coupling is not very small,
therefore the higher order corrections are usually large. On the
other hand, in the processes of top quark decays into
$\bar{B}_c(\Upsilon)$, the $t\rightarrow b\overline{c}(\bar{b}) +
c(b) + W^{+}$, there exist large scale uncertainties in the tree
level calculation \cite{Bc4}. The NLO corrections should in
principle minimize it and give a more precise prediction. To
calculate the $\bar{B}_c$ and $\Upsilon$ production rates in top
quark decays at the NLO accuracy are the aims of this work. In our
calculation, both of the S-wave spin-singlet and -triplet states are
taken into account, i.e., ${\bar{B}_{c}}^{*}$, $\bar{B}_{c}$,
$\Upsilon$ and $\eta_{b}$. To deal with the non-perturbative
effects, the non-relativistic QCD (NRQCD) \cite{Bc5} effective
theory is employed. The calculation will be performed at the NLO in
pQCD expansion, but at leading order in relativistic expansion, that
is in the expansion of $v$, the relativistic velocity of heavy
quarks inside bound states.

The paper is organized as follows: after the Introduction, in
section II we explain the calculation of leading order decay width.
In section III, virtual and real QCD corrections to Born level
result are evaluated. In section IV, the numerical calculation for
concerned processes at NLO accuracy of pQCD is performed, and the
scale dependence of the results is shown. The last section is
remained for a brief summary and conclusions.

\section{Calculation of The Born Level Decay Width}
At the leading order in $\alpha_{s}$, there are two Feynman Diagrams
for each meson production, which are shown in Figure 1. For the
convenience of analytical calculation, taking $\bar{B}_c$ as an
example, the momentum of each particle is assigned as:
$p_{1}=p_{t}$, $p_{3}=p_{b}$, $p_{4}=p_{\overline{c}}$,
$p_{5}=p_{c}$, $p_{6} = p_{W^+}$, $p_{0} = p_{3} + p_{4}$, $p_3 =
\frac{m_b}{m_c} p_4$. For bottomonium, the only difference is that
$p_{4}$ and $p_{5}$ represent the momenta of anti-bottom quark and
bottom quark which are produced in gluon splitting.

Of the $\bar{B}_c$ and $\bar{B}_c^*$ production in top quark decays,
i.e.
\begin{eqnarray}
t(p_{1})\rightarrow\bar{B}_c/{\bar{B}_c}^*(p_{0})+c(p_{5})+W^{+}(p_{6})\;
,\label{eq:1}
\end{eqnarray}
we employ the following commonly used projection operators for
quarks hadronization:
\begin{eqnarray}
v(p_{4})\,\overline{u}(p_3)& \longrightarrow& {1\over 2 \sqrt{2}}
i\gamma_5(\not\!p_{0}+m_b+m_c)\, \times \left( {1\over
\sqrt{\frac{m_b+m_c}{2}}} \psi_{\bar{B}_c}(0)\right) \otimes \left(
{{\bf 1}_c\over \sqrt{N_c}}\right)\label{eq:2}
\end{eqnarray}
and
\begin{eqnarray}
v(p_{4})\,\overline{u}(p_3)& \longrightarrow& {1\over 2 \sqrt{2}}
\not\! \epsilon_{{\bar{B}_c}^{*}}(\not\!p_{0}+m_b+m_c)\, \times
\left( {1\over \sqrt{\frac{m_b+m_c}{2}}}
\psi_{{\bar{B}_c}^*}(0)\right) \otimes \left( {{\bf 1}_c\over
\sqrt{N_c}}\right)\, .\label{eq:3}
\end{eqnarray}
Here, $\varepsilon_{{\bar{B}_c}^*}$ is the polarization vector of
${\bar{B}_c}^{*}$ with $p_{0}\cdot \varepsilon=0$, ${\bf 1}_c$
stands for the unit color matrix, and $N_c=3$ for QCD. The
nonperturbative parameters $\psi_{\bar{B}_c}(0)$ and
$\psi_{{\bar{B}_c}^{*}}(0)$ are the Schr\"{o}dinger wave functions
at the origin of $b\bar{c}$ bound states, and in the
non-relativistic limit
$\psi_{\bar{B}_c}(0)=\psi_{{\bar{B}_c}^{*}}(0)$. In our calculation,
the non-relativistic relation $m_{\bar{B}_c} =
m_{\bar{B}_c^*}=m_b+m_c$ is also adopted.
%
%
\begin{figure}[t,m,u]
\centering
\includegraphics[width=15cm,height=6cm]{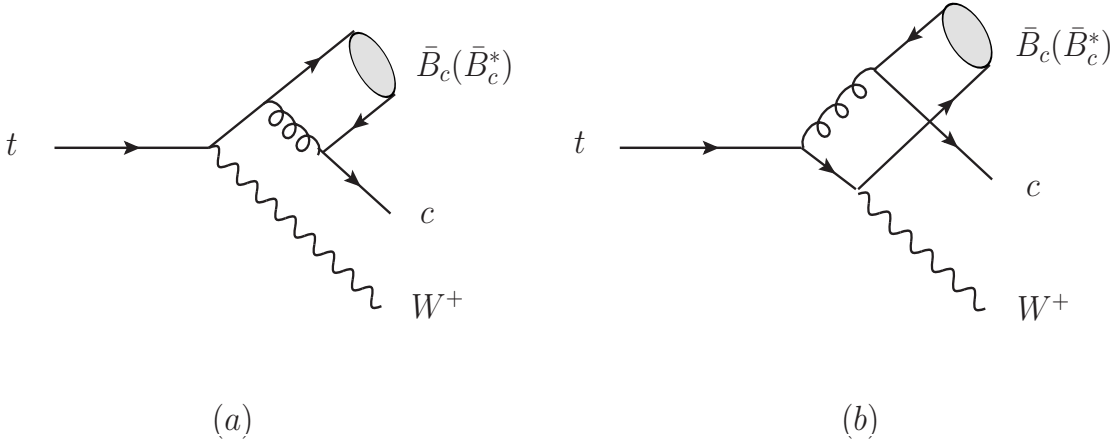}%
\caption{\small The leading order Feynman diagrams for $\bar{B}_c$
and $\bar{B}_c^{*}$ production in top quark decays.} \label{graph1}
\end{figure}

The LO amplitudes for $\bar{B}_c$ production can then be readily
obtained with above preparations. They are:
\begin{eqnarray}
{\cal M}_{a}=\frac{\pi\alpha_{s}g\psi_{\bar{B}_c}(0)V_{tb}C_{F}
\delta_{j,k}}{\sqrt{6m_{\bar{B}_c}}}
\bar{u}(p_{5})\gamma^{\mu}i\gamma_{5}(\not\!p_{0}+m_{\bar{B}_c})
\gamma_{\mu}\frac{(\not\!p_{0}
+\not\!p_{5}+m_{b})}{(p_{0}+p_{5})^{2}-m_{b}^{2}}
\frac{\not\!\epsilon(p_{6})(1-\gamma_{5})}{({p_{4}+p_{5}})^{2}}u(p_{1})\;
, \label{eq:4}
\end{eqnarray}
and
\begin{eqnarray}
{\cal M}_{b}=\frac{\pi\alpha_{s}g\psi_{\bar{B}_c}(0)V_{tb}C_{F}
\delta_{j,k}}{\sqrt{6m_{\bar{B}_c}}}
\bar{u}(p_{5})\gamma^{\mu}i\gamma_{5}(\not\!p_{0}+m_{\bar{B}_c})
\frac{\not\!\epsilon(p_{6})
(1-\gamma_{5})}{({p_{4}+p_{5}})^{2}}\frac{(\not\!p_{3}
+\not\!p_{6}+m_{t})}{(p_{3}+p_{6})^{2}-m_{t}^{2}}\gamma_{\mu}u(p_{1})\;
.\label{eq:5}
\end{eqnarray}
Here, $j$, $k$ are color indices, $C_F=4/3$ belongs to the $SU(3)$
color structure. For ${\bar{B}_c}^*$ production, the amplitudes can
be obtained by simply substituting $i\gamma_{5}(\not\!\!p_{0} +
m_{\bar{B}_c})$ with $\not\!\epsilon_{{\bar{B}_c}^{*}}
(\not\!p_{0}+m_b+m_c)$ in above expressions.

The Born amplitude of the processes shown in Fig.1 is then ${\cal
M}_{Born} = {\cal M}_{a} + {\cal M}_{b}$, and subsequently, the
decay width at leading order reads:
\begin{eqnarray}
\mathrm{d}\Gamma_{Born}=\frac{1}{2m_{t}}\frac{1}{2}\frac{1}{N_c}
\sum|{\cal M}_{Born}|^{2}\mathrm{d}\textmd{PS}_{3}\; .\label{eq:6}
\end{eqnarray}
Here, $\sum$ represents the sum over polarizations and colors of the
initial and final particles, $\frac{1}{2}$ and $\frac{1}{N_c}$ come
from spin and color average of initial t quark,
$\mathrm{d}\textmd{PS}_{3}$ stands for the integrants of three-body
phase space, whose concrete form is
\begin{eqnarray}
\mathrm{d}\textmd{PS}_{3}=\frac{1}{32\pi^3}\frac{1}{4m_t^2}
\mathrm{d}s_{1}\mathrm{d}s_{2}\; ,\label{eq:7}
\end{eqnarray}
where $s_{1}=(p_{0}+p_{5})^{2}=(p_{1}-p_{6})^{2}$ and $s_{2}=
(p_{5}+p_{6})^{2}=(p_{1}-p_{0})^{2}$ are Mandelstam variables. The
upper and lower bounds of the above integration
are
\begin{eqnarray}
s_{1}^{max}=&&\frac{\sqrt{f[m_t^2,s_2,m_{\bar{B}_c}^2]\cdot
f[s_2,m_c^2,m^2_{W}]}+[m_t^2-s_2-(m_b+m_c)^2](s_2+m_c^2-m^2_{W})}
{2s_2}\nonumber\\
&& +m_{\bar{B}_c}^2+m_c^2\; ,\label{eq:9}
\end{eqnarray}

\begin{eqnarray}
s_{1}^{min}=&&-\frac{\sqrt{f[m_t^2,s_2,m_{\bar{B}_c}^2]\cdot
f[s_2,m_c^2,m^2_{W}]}-[m_t^2-s_2-(m_b+m_c)^2](s_2+m_c^2-m^2_{W})}
{2s_2}\nonumber\\
&& +m_{\bar{B}_c}^2+m_c^2 \label{eq:10}
\end{eqnarray}
and
\begin{eqnarray}
s_{2}^{max} = [m_{t}-(m_{b}+m_{c})]^{2}\; ,\;  s_{2}^{min} =
(m_{c}+m_{W})^{2}
\end{eqnarray}
with
\begin{eqnarray}
f[x,y,z]&=&(x-y-z)^2-4yz\; .
\end{eqnarray}

\section{The Next-to-Leading Order Corrections}
\begin{figure}[t,m,u]
\centering
\includegraphics[width=15cm,height=6cm]{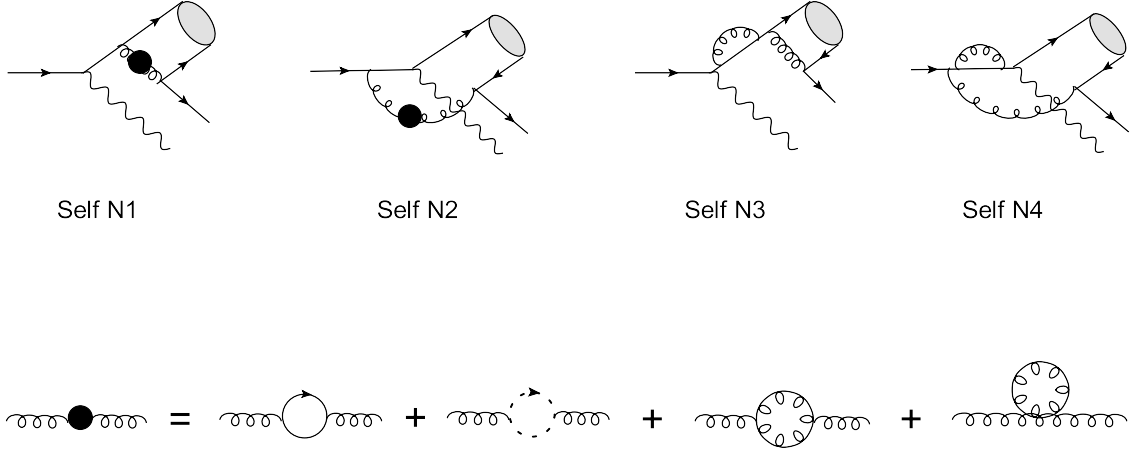}%
\caption{\small The self-energy diagrams in virtual corrections.}
\label{graph2}
\end{figure}

\begin{figure}[t,m,u]
\centering
\includegraphics[width=16cm,height=12cm]{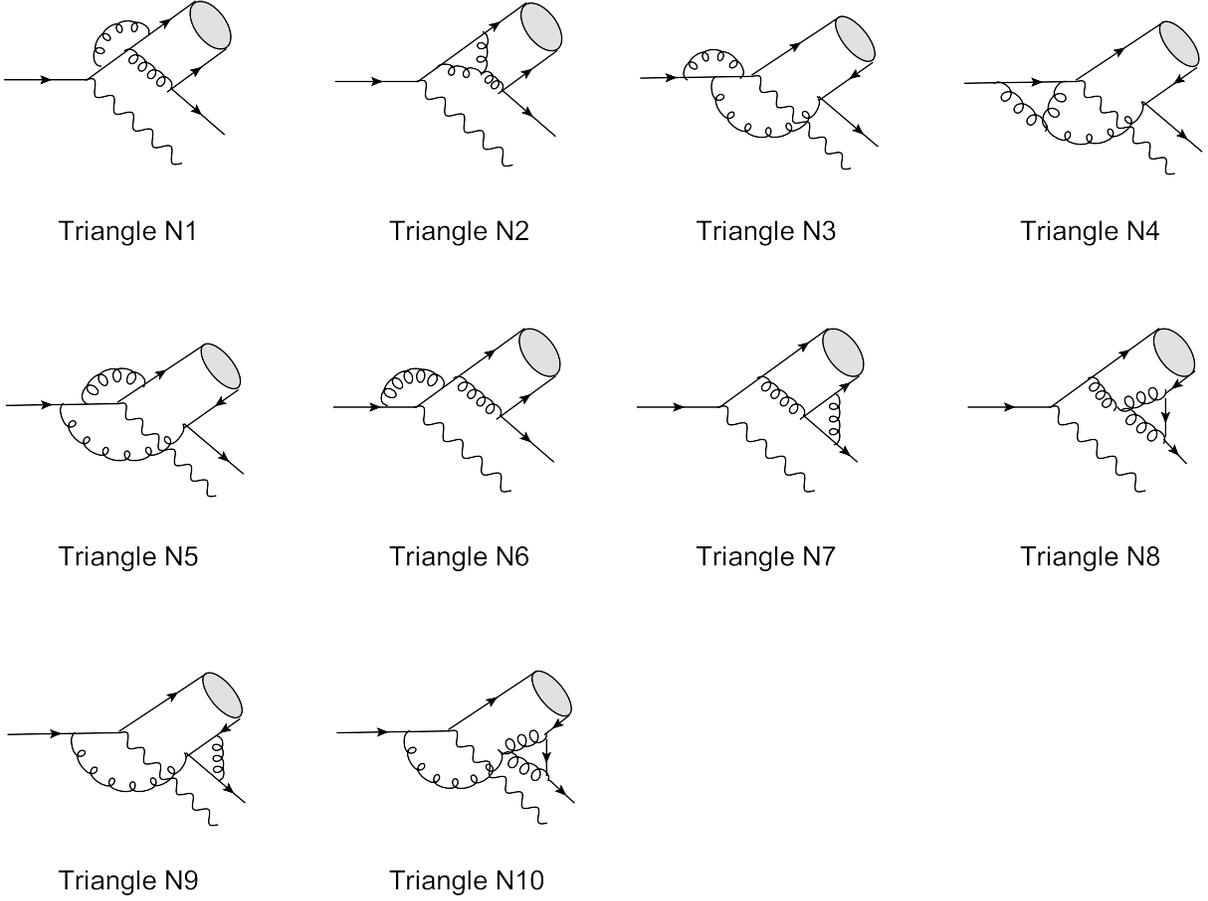}%
\caption{\small The triangle diagrams in virtual corrections.}
\label{graph3}
\end{figure}

\begin{figure}[t,m,u]
\centering
\includegraphics[width=15cm,height=11cm]{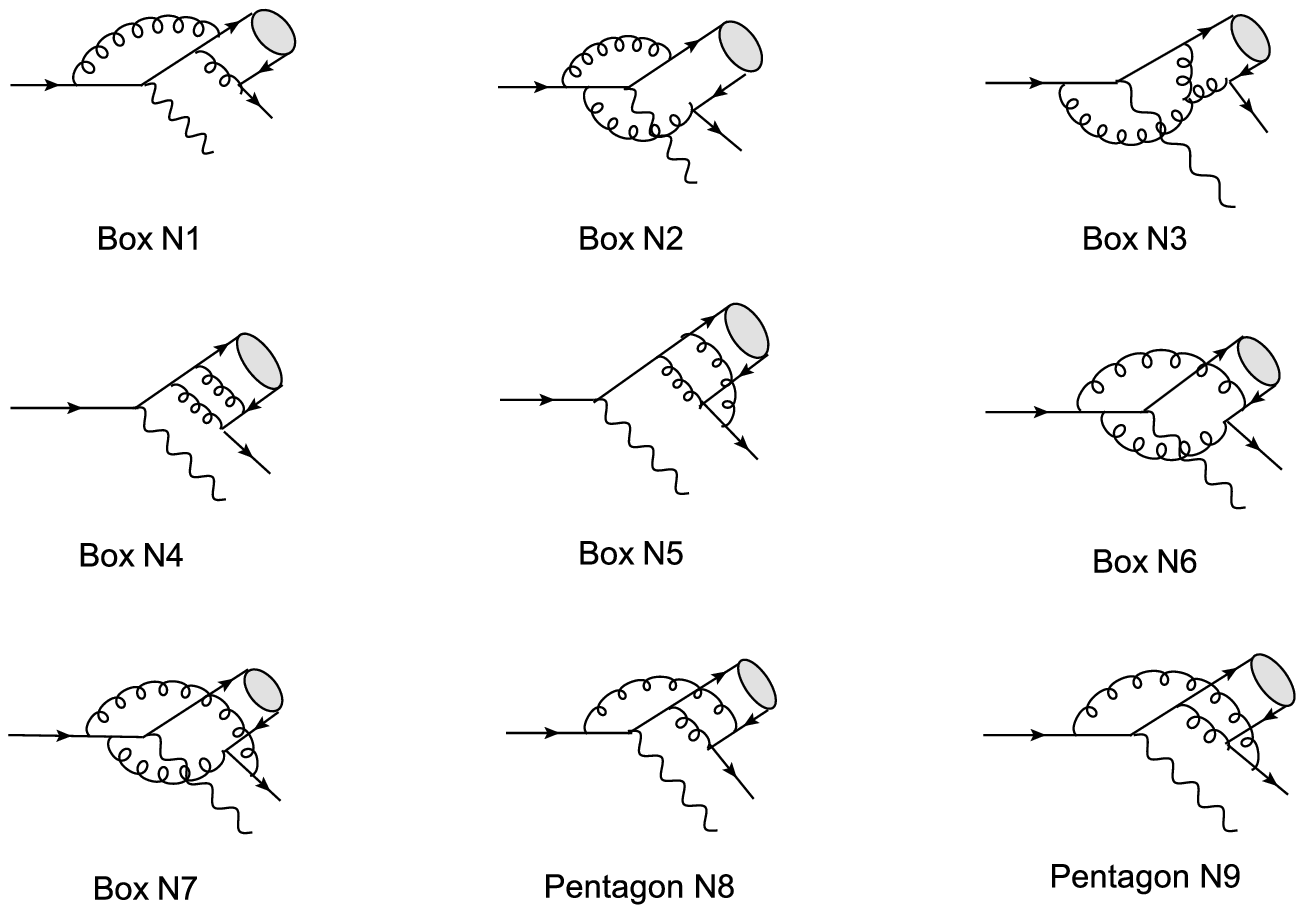}%
\caption{\small The box and pentagon diagrams in virtual
corrections.} \label{graph4}
\end{figure}

\begin{figure}[t,m,u]
\centering
\includegraphics[width=15cm,height=9cm]{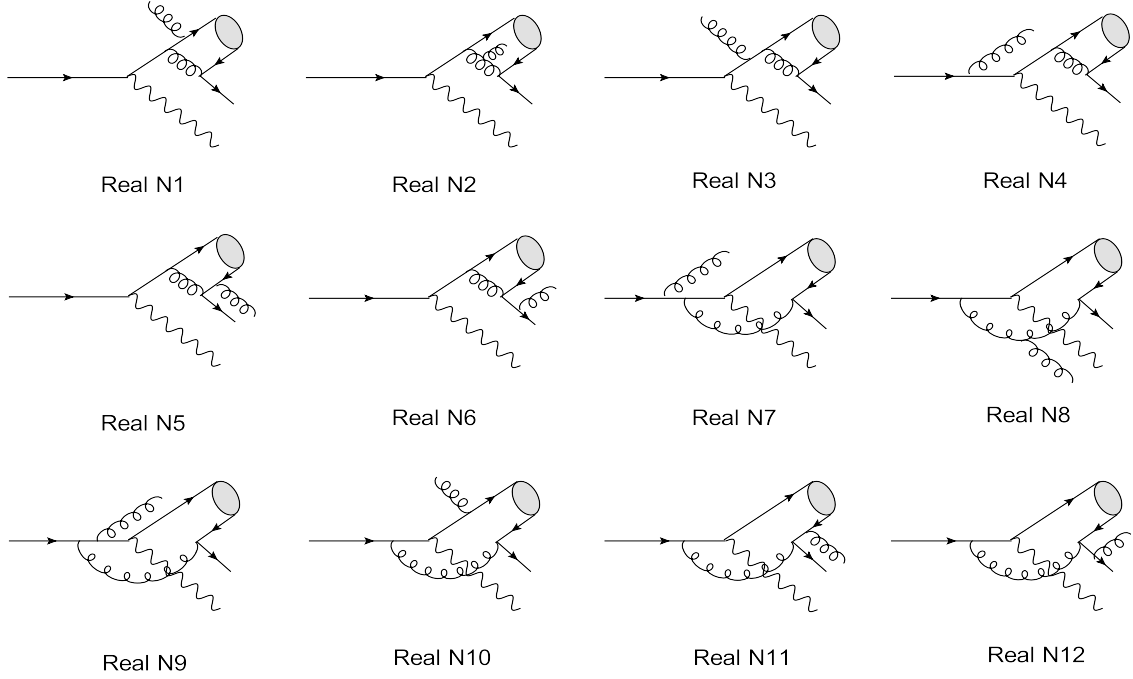}%
\caption{\small The real correction Feynman diagrams that contribute
to the production of $\bar{B}_c$ or $\bar{B}_c^{*}$.} \label{graph5}
\end{figure}

At the next-to-leading order, the top quark decays to $\bar{B}_c$
and $\Upsilon$ include the virtual and real QCD corrections to the
leading order process, as shown in Figs.\ref{graph2}-\ref{graph5}.
With virtual corrections, the decay widths at the NLO can be
formulated as
\begin{eqnarray}
\mathrm{d}\Gamma_{Virtual}=\frac{1}{2m_{t}}\frac{1}{2}
\frac{1}{N_c}\sum2\textmd{Re} ({\cal M}_{Born}^{*} {\cal
M}_{Virtual}) \mathrm{d}\textmd{PS}_{3}\; .\label{eq:12}
\end{eqnarray}
The ultraviolet(UV) and infrared(IR) divergences usually exist in
virtual corrections. We use the dimensional regularization scheme to
regularize the UV and IR divergences, similar as performed in
Ref.\cite{Bc6}, and the Coulomb divergence is regularized by the
relative velocity $v$. In dimensional regularization, $\gamma_{5}$
is difficult to deal with. In this calculation, we adopt the Naive
scheme, that is, $\gamma_{5}$ anticommutates with each $\gamma^\mu$
matrix in d-dimension space-time, $\{\gamma_{5},\gamma^{\mu}\}=0$.
The UV divergences exist merely in self-energy and triangle
diagrams, which can be renormalized by counter terms. The
renormalization constants include $Z_{2}$, $Z_{3}$, $Z_{m}$, and
$Z_{g}$, corresponding to quark field, gluon field, quark mass, and
strong coupling constant $\alpha_{s}$, respectively. Here, in our
calculation the $Z_{g}$ is defined in the
modified-minimal-subtraction ($\mathrm{\overline{MS}}$) scheme,
while for the other three the on-shell ($\mathrm{OS}$) scheme is
adopted, which tells
\begin{eqnarray}
&&\hspace{-0.3cm}\delta Z_m^{OS}=-3C_F
\frac{\alpha_s}{4\pi}\left[\frac{1}{\epsilon_{UV}}-\gamma_{E}+
\ln\frac{4\pi\mu^{2}}{m^{2}}+\frac{4}{3}+{\mathcal{O}}(\epsilon)\right]\;
,\nonumber
\\ &&\hspace{-0.3cm}\delta Z_2^{OS}=-C_F
\frac{\alpha_s}{4\pi}\left[\frac{1}
{\epsilon_{UV}}+\frac{2}{\epsilon_{IR}}-3\gamma_{E}+3\ln
\frac{4\pi\mu^{2}}{m^{2}}+4+{\mathcal{O}}(\epsilon)\right]\; ,\nonumber\\
&&\hspace{-0.3cm}\delta Z_3^{OS}= \frac{\alpha_s}{4\pi}
\left[(\beta_0-2C_A)(\frac{1}{\epsilon_{UV}}-
\frac{1}{\epsilon_{IR}})+{\mathcal{O}}(\epsilon)\right]\; ,\nonumber\\
&&\hspace{-0.3cm}\delta Z_g^{\overline{MS}}=-\frac{\beta_0}{2}
\frac{\alpha_s}{4\pi}\left[\frac{1}{\epsilon_{UV}}-\gamma_{E}+
\ln4\pi+{\mathcal{O}}(\epsilon)\right]\; .\label{eq:13}
\end{eqnarray}
Here, $\beta_{0}=(11/3)C_{A}-(4/3)T_{f}n_{f}$ is the one-loop
coefficient of the QCD beta function; $n_{f}=5$ is the number of
active quarks in our calculation; $C_{A}=3$ and $T_{F}=1/2$
attribute to the SU(3) group; $\mu$ is the renormalization scale.

In virtual corrections, IR divergences remain in the triangle and
box diagrams. Of all the triangle diagrams, only two have IR
divergences, which are denoted as $\mathrm{Triangle N7}$ and
$\mathrm{Triangle N9}$ in Fig.\ref{graph3}. Of the diagrams in
Fig.\ref{graph4}, $\mathrm{Box N3}$ has no IR singularity, while
$\mathrm{Box N4}$ and $\mathrm{Pentagon N9}$ have Coulomb
singularities and $\mathrm{Pentagon N9}$ possesses ordinary IR
singularity as well. The remaining diagrams all have IR
singularities, while the combinations $\mathrm{Box N2+Box N6}$,
$\mathrm{Box N1+Pentagon N8+Triangle N9}$, $\mathrm{Box N5+Triangle
N7}$ are IR finite. The Coulomb singularities belonging to
$\mathrm{Box N4}$ and $\mathrm{Pentagon N9}$ can be regularized by
the relative velocity $v$. After regularization procedure, the
$\frac{1}{\epsilon}$ term will be canceled out by the counter terms
of external quarks which form the $\bar{B}_c$ or $\Upsilon$, while
the $\frac{1}{v}$ term will be mapped onto the wave functions of the
concerned heavy mesons. The remaining IR singularities in
$\mathrm{Box N7}$ and $\mathrm{Box N9}$ are canceled by the
corresponding parts in real corrections. In the end, the IR and
Coulomb divergences in virtual corrections can be expressed as
\begin{eqnarray}
\mathrm{d}\Gamma_{virtual} ^{IR,Coulomb}=
\mathrm{d}\Gamma_{Born}\frac{ 4 \alpha_s}{3\pi} \left[
\frac{\pi^2}{v}-\frac{1}{ \epsilon}- \frac{p_t \cdot p_{c}x_{s} \ln
x_{s}}{m_{c}m_{t}(1-x_{s}^{2})} \frac{1}{\epsilon} \right]\;
,\label{eq:14}
\end{eqnarray}
with $p_{t}=p_{1}$, $p_{c}=p_5$ and $x_{s} =
\frac{1-\sqrt{1-2m_{c}m_{t}/(m_{c}m_{t}-p_t \cdot p_{c})}} {1 +
\sqrt{1-2m_{c}m_{t}/(m_{c}m_{t}-p_t \cdot p_{c})}}$. Here, in this
work $\frac{1}{\epsilon}$ in fact represents
$\frac{1}{\epsilon}-\gamma_{E}+\ln(4\pi\mu^{2})$.

Of our concerned processes, there are $12$ different diagrams in
real correction, as shown in Fig.\ref{graph5}. Among them,
$\mathrm{Real N2}$, $\mathrm{Real N3}$, $\mathrm{Real N8}$, and
$\mathrm{Real N9}$ are IR-finite, meanwhile the combinations of
$\mathrm{Real N1+ Real N5}$ and $\mathrm{Real N10+Real N11}$ exibit
no IR singularities as well, due to the reasons of gluon connecting
to the $b$ or $\bar{c}$ quark of final $\bar{B}_c$ or $\Upsilon$.
The remaining diagrams, $\mathrm{Real N4}$, $\mathrm{Real N6}$,
$\mathrm{Real N7}$, and $\mathrm{Real N12}$ are not IR singularity
free. To regularize the IR divergence, we enforce a cut on the gluon
momentum, the $p_{7}$. The gluon with energy $p_{7}^{0} < \delta$ is
considered to be soft, while $p_{7}^{0} > \delta$ is thought to be
hard. The $\delta$ is a small quantity with energy-momentum unit. In
this case, the IR term of the decay width can then be written as:
\begin{eqnarray}
 \mathrm{d}\Gamma_{Real}^{IR}=\frac{1}{2m_{t}}\frac{1}{2}\frac{1}{N_c}
 \sum|{\cal M}_{Real}|^{2}\; \mathrm{d}
 \textmd{PS}_{4}\mid_{soft}\; ,\label{eq:15}
\end{eqnarray}
where $\mathrm{d}\textmd{PS}_4$ is the four-body phase space
integrants for real correction. Under the condition of
$p_{7}^{0}<\delta$, in the Eikonal approximation we obtain
\begin{eqnarray}
\mathrm{d}\textmd{PS}_{4}\mid_{soft}=\mathrm{d}\textmd{PS}_{3}
\frac{d^{3}p_{7}}{(2\pi)^{3}2p_{7}^{0}}\mid_{p_{7}^{0}<\delta}\;
\end{eqnarray}
In the small $\delta$ limit, the IR divergent terms in real
correction can therefore be expressed as
\begin{eqnarray}
\mathrm{d}\Gamma_{Real}^{IR}=\mathrm{d}\Gamma_{Born}
\frac{4\alpha_s}{3 \pi}
\left\{\left(\frac{1}{\epsilon}-\textmd{Log}(\delta^2)\right)\left[1+
\frac{p_t \cdot p_{c}x_{s} \ln
x_{s}}{m_{c}m_{t}(1-x_{s}^{2})}\right] + \rm{finite\;
terms}\right\}\; . \label{eq:17}
\end{eqnarray}
Here, the $\textmd{Log}(\delta^2)$ involved terms will be canceled
out by the $\delta$-dependent terms in the hard sector of real
corrections. Referring to the Eq.(\ref{eq:14}), it is obvious that
the IR divergent terms in real and virtual corrections cancel with
each other. In case of hard gluons in real correction, the decay
width reads
\begin{eqnarray}
 \mathrm{d}\Gamma_{Real}^{hard}=\frac{1}{2m_{t}}\frac{1}{2}\frac{1}{N_c}
 \sum|{\cal M}_{Real}|^{2}\; \mathrm{d}
 \textmd{PS}_{4}\mid_{hard}\; .\label{eq:18}
\end{eqnarray}
In this case the phase space $\mathrm{d}\textmd{PS}_{4}\mid_{hard}$
can be written as
\begin{eqnarray}
     \mathrm{d}\textmd{PS}_{4}\mid_{hard}=\frac{2}{(4\pi)^{6}}&&
     \frac{\sqrt{(sy+m_{c}^{2}-m_{W}^{2})^{2}-4sym_{c}^{2}}}{y}
      \int_{{p_{0}}^{0}_{-}}^{{p_{0}}^{0}_{+}}\mathrm{d}{p_{0}}^{0}
      \int_{-1}^{1}\mathrm{d}\cos\theta_{c}\int_{0}^{2\pi}\mathrm{d}
      \phi_{c}\nonumber\\
      &&\times\left\{\int_{\delta}^{{p_{7}}^{0}_{-}}\mathrm{d}{p_{7}}^{0}
      \int_{y_{-}}^{y_{+}}\mathrm{d}y+
      \int_{{p_{7}}^{0}_{-}}^{{p_{7}}^{0}_{+}}\mathrm{d}{p_{7}}^{0}
      \int_{\frac{(m_{c}+m_{W})^{2}}{s}}^{y_{+}}\mathrm{d}y\right\}
      \label{eq:20}
\end{eqnarray}
with
\begin{eqnarray}
&&{p_{0}}^{0}_{-}=m_{b}+m_{c}\; ,\\
&&{p_{0}}^{0}_{+}=\frac{s+m_{b}^{2}-m_{W}^{2}+2m_{b}m_{c}-2m_{W}
\cdot m_{c}}{2\sqrt{s}}\; , \\
&&{p_{7}}^{0}_{-} =
\frac{s+m_{b}^{2}-m_{W}^{2}+2m_{b}m_{c}-2m_{W}\cdot
m_{c}-2\sqrt{s}{p_{0}}^{0}} {2\sqrt{s}-2{p_{0}}^{0} + 2
\sqrt{|\overrightarrow{{p_{0}}}|}}\; ,\\
&&{p_{7}}^{0}_{+}=\frac{s+m_{b}^{2}-m_{W}^{2}+2m_{b}m_{c}-2m_{W}\cdot
m_{c}-2\sqrt{s}{p_{0}}^{0}} {2\sqrt{s}-2{p_{0}}^{0} -
2\sqrt{|\overrightarrow{p_{0}}|}}\; ,\\
&&y_{-}=\frac{1}{s}[(\sqrt{s}-{p_{0}}^{0}-{p_{7}}^{0})^{2}-|
\overrightarrow{{p_{0}}}|^{2}-({p_{7}}^{0})^{2}
  -2|\overrightarrow{p_{0}}|p_{7}^{0}]\; ,\\
&&y_{+}=\frac{1}{s}[(\sqrt{s}-{p_{0}}^{0}-{p_{7}}^{0})^{2}-|
\overrightarrow{{p_{0}}}|^{2}-({p_{7}}^{0})^{2}
  +2|\overrightarrow{p_{0}}|{p_{7}}^{0}]\; ,
\end{eqnarray}
where $y$ is a dimensionless parameter defined as $y =
(p_1-p_0-p_7)^2/s$ with $\sqrt{s}=m_{t}$, and
\begin{eqnarray}
|\overrightarrow{p_{0}}|=\sqrt{({p_0}^0)^{2}-m_{\bar{B}_{c}}^{2}}\;
.~~~~~~~~~~~~~~~~~~~~~~~~~~~~~~~~~~~\label{eq:21}
\end{eqnarray}
The sum of the soft and hard sectors gives the total contribution of
real corrections, i.e., $\Gamma_{Real} = \Gamma_{Real}^{IR} +
\Gamma_{Real}^{hard}$.

With the real and virtual corrections, we then obtain the total
decay width of $t$ quark to $\bar{B}_c$ and $\Upsilon$ at the NLO
accuracy of QCD
\begin{eqnarray}
\Gamma_{total}=\Gamma_{Born}+\Gamma_{Virtual}+\Gamma_{Real}+
{\mathcal{O}}(\alpha_s^4)\; .\label{eq:22}
\end{eqnarray}
In above expression, the decay width is UV and IR finite. In our
calculation the FeynArts \cite{feynarts} was used to generate the
Feynman diagrams, the amplitudes were generated by the FeynCalc
\cite{feyncalc}, and the LoopTools \cite{looptools} was employed to
calculate the Passarino-Veltman integrations. The numerical
integrations of the phase space were performed by the MATHEMATICA.

\section{Numerical results}

To complete the numerical calculation, the following ordinarily
accepted input parameters are taken into account:
\begin{eqnarray}
&&m_{b}=4.9\; \textmd{GeV},\; m_{c}=1.5\; \textmd{GeV},\;
m_{t}=174\; \textmd{GeV},\; m_{W}=80\; \textmd{GeV}\; ,
\end{eqnarray}
\begin{eqnarray}
\psi_{\bar{B}_c}(0)&=&\psi_{\bar{B}_c^*}(0)=\frac{R_{1}(0)}
{\sqrt{4\pi}}=0.3616\; \mathrm{GeV}^{3/2}\; ,\\
\psi_{\Upsilon}^{LO}(0)&=&\psi_{\eta_{b}}^{LO}(0)=
\frac{R_{2}^{LO}(0)}{\sqrt{4\pi}}=0.6812\; \mathrm{GeV}^{3/2}\; ,\\
\psi_{\Upsilon}^{NLO}(0)&=&\psi_{\eta_{b}}^{NLO}(0)=
\frac{R_{2}^{NLO}(0)}{\sqrt{4\pi}}=\frac{R_{2}^{LO}(0)}
{\sqrt{4\pi-16C_F\alpha_s}}=0.8277\; \mathrm{GeV}^{3/2}\; ,
\end{eqnarray}
\begin{eqnarray}
V_{tb}=1.0,\;\;\;\; G_F=1.1660\times 10^{-5}\; \mathrm{GeV}^{-2} \;
.\label{eq:23}
\end{eqnarray}
Here, $V_{tb}$ is the Cabibbo-Kobayashi-Maskawa(CKM) matrix element
and $G_F$ is weak interaction Fermi constant.

In above numerical calculation inputs, the radial wave function at
the origin for S-wave $\bar{B}_c^*(\bar{B}_c)$ system is estimated
by potential model \cite{Bc8}, while the corresponding
$\Upsilon(\eta_b)$ nonperturbative parameter is determined from its
electronic decay rate \cite{Bc7}. One loop result of strong coupling
constant is taken into account, i.e.
\begin{eqnarray}
\alpha_s(\mu)=\frac{4\pi}{(11-\frac{2}{3}n_f)\mathrm{Log}(\frac{\mu^2}
              {\Lambda_{QCD}^2})}\; .\label{eq:24}
\end{eqnarray}

\begin{figure}
\centering
\includegraphics[width=0.480\textwidth]{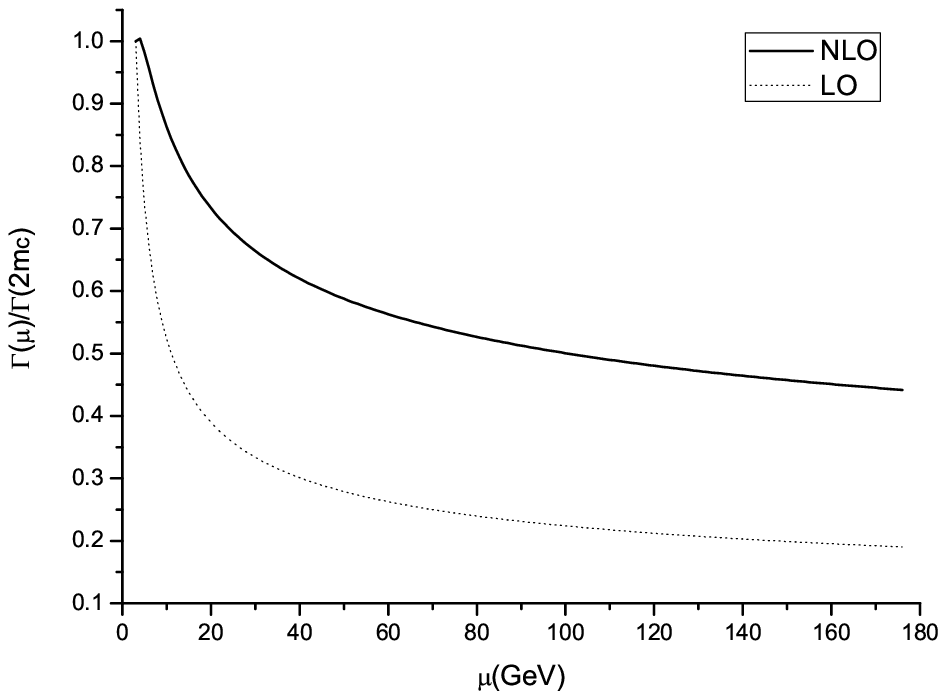}%
\hspace{5mm}
\includegraphics[width=0.480\textwidth]{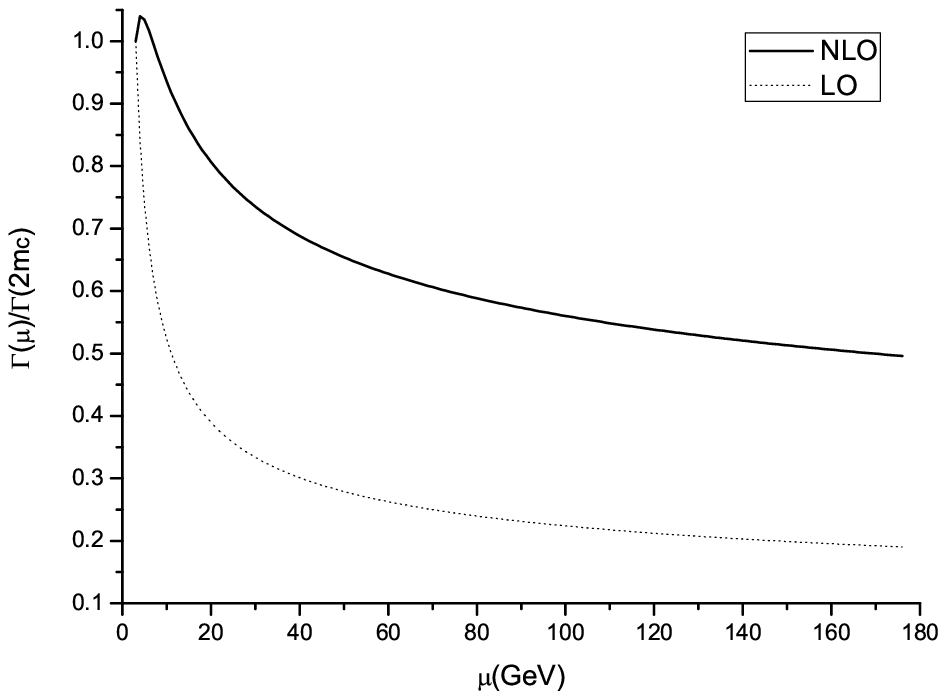}\hspace*{\fill}
\caption{\small The ratio $\Gamma(\mu)/\Gamma(2m_c)$ versus
renormalization scale $\mu$ in $t$ quark decays. The left diagram
for the $b\bar{c}$ spin-singlet state $\bar{B}_c$ and the right
diagram for the spin-triplet state $\bar{B}_c^*$.} \label{lpty1}
\vspace{-0mm}
\end{figure}

\begin{figure}
\centering
\includegraphics[width=0.480\textwidth]{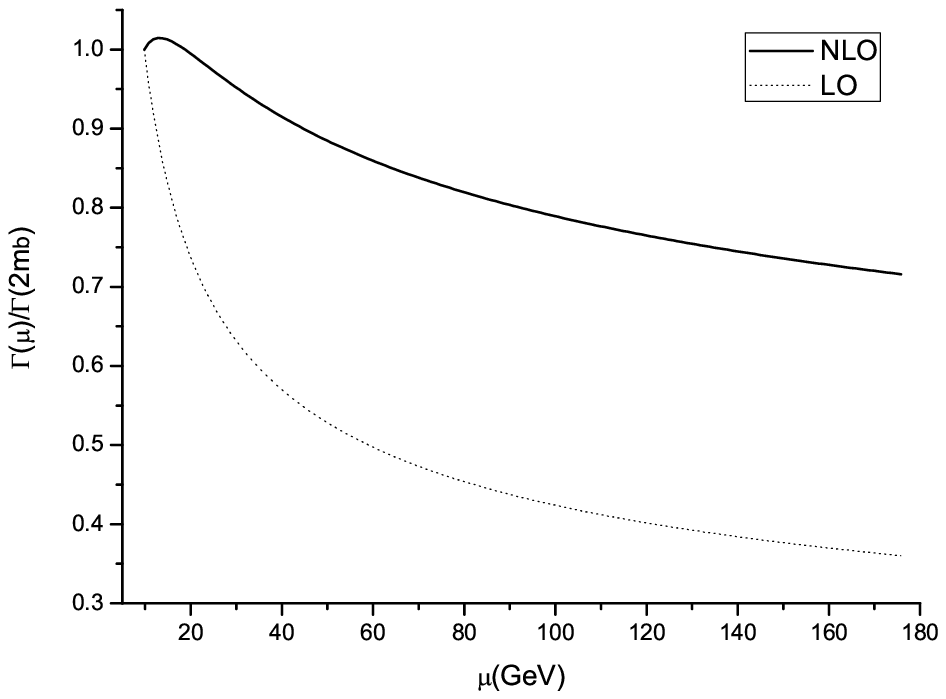}%
\hspace{5mm}
\includegraphics[width=0.480\textwidth]{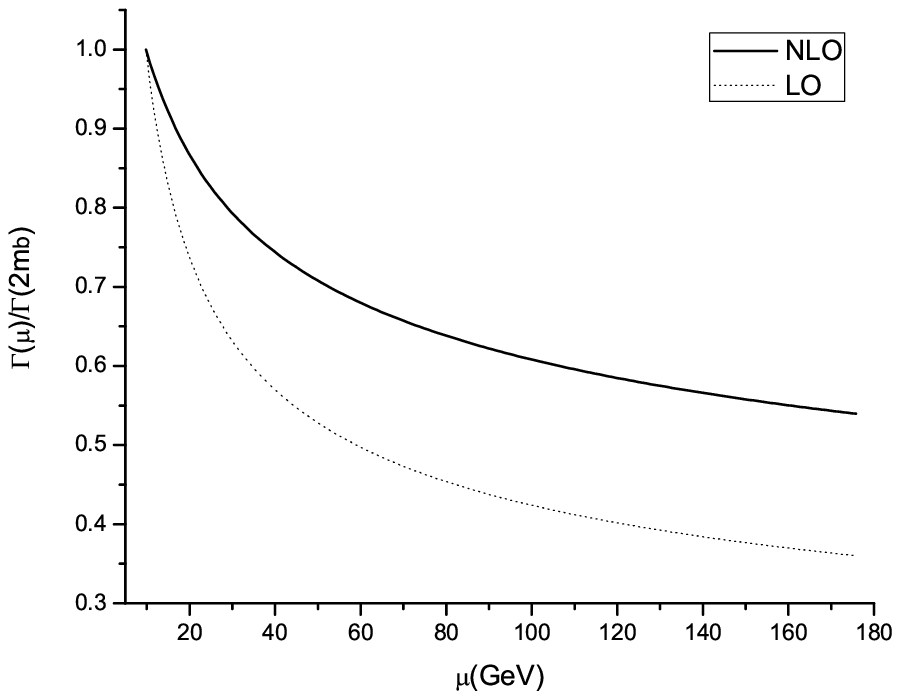}\hspace*{\fill}
\caption{\small The ratio $\Gamma(\mu)/\Gamma(2m_b)$ versus
renormalization scale $\mu$ in $t$ quark decays. The left diagram
for the $b\bar{b}$ spin-singlet state $\eta_{b}$ and the right
diagram for the spin-triplet state $\Upsilon$.} \label{lpty2}
\vspace{-0mm}
\end{figure}

With the above preparation, one can readily obtain the decay widths
of top quark to $b\bar{c}$ and $b\bar{b}$ mesons, as listed in Table
\ref{ratio2}. To see the scale dependence of the LO and NLO results,
the ratios $\Gamma(\mu)/\Gamma(2m_{c})$ for $b\bar{c}$ system and
$\Gamma(\mu)/\Gamma(2m_{b})$ for $b\bar{b}$ system are showed in
Figures \ref{lpty1} and \ref{lpty2}, respectively. Calculation tells
that after including the NLO corrections, the energy scale
dependence of the results is reduced, as expected.

\begin{table}
\begin{center}
\caption{The decay widths of the processes
$t\rightarrow\bar{B}_c^*+W^++c$, $t\rightarrow\bar{B}_c+W^++c$,
$t\rightarrow\Upsilon+W^++b$ and $t\rightarrow\eta_b+W^++b$ at the
tree level and with the NLO QCD corrections are presented in two
renormalization scale $\mu$ limits, those are $2m_c$ and $m_t$ for
the first two processes and $2m_b$ and $m_t$ for the other two.}
\vspace{1mm}
\begin{tabular}{|c||c|c|c|c|c|c|c|c|}
\hline\hline
&\multicolumn{2}{|c|}{$t\rightarrow\bar{B}_c^*+W^++c$}&
\multicolumn{2}{|c|}{$t\rightarrow\bar{B}_c+W^++c$}
&\multicolumn{2}{|c|}{$t\rightarrow\Upsilon+W^++b$}&
\multicolumn{2}{|c|}{$t\rightarrow\eta_b+W^++b$}\\
\hline\hline $\mu$ &~$2m_c$~ & ~$m_t$~ & ~$2m_c$~
& ~$m_t$~ & ~$2m_b$~ &~$m_t$~ & ~$2m_b$~ & ~$m_t$~ \\
\hline\hline $\Gamma_{LO}$ & 0.793MeV & 0.151MeV & 0.572MeV &
0.109MeV & 26.8keV & 9.54keV & 27.1keV &
9.67keV \\
\hline\hline $\Gamma_{NLO}$ & 0.619MeV & 0.307MeV & 0.514MeV &
0.227MeV & 52.3keV & 28.2keV & 34.3keV & 24.5keV\label{tb1}
\\
\hline\hline
\end{tabular}
\label{ratio2}
\end{center}
\end{table}

\section{Summary and Conclusions}
In this work we have calculated the decay widths of top quark to
S-wave $b\bar{c}$ and $b\bar{b}$ bound states at the NLO accuracy of
perturbative QCD. Considering that there will be copious $t\bar{t}$
data in the near future at the LHC, our results are helpful to the
study of the indirect production of these states. They may be also
useful to the future study of NLO heavy quark to $b\bar{c}$ and
$b\bar{b}$ bound states fragmentation functions.

Numerical results indicate that the NLO corrections greatly enhance
the LO results for $b\bar{b}$ system, while slightly decrease the
$b\bar{c}$ states production widths. The main reason for this
difference is that the NLO wave function for bottomonium is much
larger than that of LO one, while for the calculation of $\bar{B}_c$
meson, the same wave function given by potential model is used.
Although from Table I, superficially the number of indirectly
produced $\bar{B}_c$ overshoots that of $\Upsilon$, experimentally
to detect the latter is much easier than the former. Since top quark
dominantly decays into $b$ and $W^+$ final state with a width of 1.5
GeV or so, numerical results remind us that the $\Upsilon$ indirect
production from top quark decay is detectable, while it is hard to
pin down the $\bar{B}_c$ states by this way.

The numerical calculation also shows that the next-to-leading order
QCD corrections to processes $t\rightarrow b\bar{c}(b\bar{b}) + W^+
+c(b)$ decrease the energy scale dependence of the decay widths as
expected, and hence the uncertainties in theoretical estimation.
Future precise experiment on the concerned processes may provide a
test on the theoretical framework for heavy quarkonium production
and the reliability of perturbative calculation for them.

\vspace{1.1cm} {\bf Acknowledgments}

This work was supported in part by the National Natural Science
Foundation of China(NSFC) under the grants 10935012, 10928510,
10821063 and 10775179, by the CAS Key Projects KJCX2-yw-N29 and
H92A0200S2.


\end{document}